\documentclass[11pt]{kluwer}
\usepackage{graphics}
\begin{opening}
\title{TRANSPORT IN PERTURBED\\
INTEGRABLE HAMILTONIAN SYSTEMS\\
AND THE FRACTALITY OF PHASE SPACE}
\runningtitle{TRANSPORT IN PERTURBED
INTEGRABLE HAMILTONIAN SYSTEMS}
\date{}
\author{H. Varvoglis, Ch. Vozikis and B. Barbanis}
\institute{Section of Astrophysics, Astronomy and Mechanics,\\
Department of Physics, Aristotle University of Thessaloniki\\
GR-54006 THESSALONIKI, GREECE}
\end{opening}

\begin{document}

\begin{abstract}
We study transport in a model perturbed integrable Hamiltonian system
by calculating the volume, $V(t)$, of elementary phase space cells visited
by a trajectory, as a function of time. We use this function in order to
"measure" the fractality of phase space. We argue that the "degree" of
fractality is related to the well known difficulties in assigning
unambiguously Lyapunov Characteristic Numbers (LCN's ) to
trajectories. Moreover we show that transport in phase space regions
with pronounced fractality cannot be described as "normal diffusion",
since the self-similar properties of $V(t)$ imply that it is governed by
Levy statistics, while the correlation dimension of $dV/dt$ implies that,
in some cases, the process is strongly non-Markovian.
\end{abstract}

\section{Introduction}
Transport in phase space is an interesting theoretical aspect of chaotic
behaviour in perturbed integrable Hamiltonian systems, since it is
encountered in many problems of physical interest (e.g. see Kaneko and
Konishi 1989, Meiss 1992, Shlesinger {\it et al.} 1993, Benkadda {\it et
al.} 1994 and references therein). Recently, in a series of papers by
Lecar {\it et al.} (e.g. see Murison {\it et al.} 1994 and references
therein), a correlation was reported between transport in action space
of the Elliptical Restricted Three Body Problem (ERTBP) and the
Lyapunov Characteristic Numbers (LCN's). In particular the authors,
using as a model the ERTBP, presented numerical evidence for the
existence of a power law, relating the exit times of asteroidal trajectories
from action space regions with the corresponding LCN's of the
trajectories. A theoretical interpretation of this correlation, in cases of
"strong" perturbation, where the motion can be considered as a random
walk in action space, has been attempted by Varvoglis and Anastasiadis
(1996) and Morbidelli and Froeschl\'e (1996).

However, in attempting a statistical description of the process, in order
to relate the local rate of trajectory divergence to transport, there are
two important problems that have to be solved. The first refers to the
difficulty in calculating reliably LCN's in cases where a trajectory is
continuously migrating to new regions of phase space, so that the usual
method of calculating LCN's (as the limit of $\chi (t) = ln(\xi / \xi
_0)/t,  t \to \infty$) does not show signs of convergence. The second
refers to the fact that, in the case of the main asteroidal belt, the
perturbation to the motion of asteroids by Jupiter cannot be considered
as "strong". Both problems originate from the fact that the invariant tori
in the regions of interest are far from being completely destroyed, so
that the phase space is characterised by a pronounced fractality. This, in
turn, implies that transport (i) obeys Levy rather than classical Brownian
kinetics (e.g. Shlesinger et al. 1993, Benkadda et al. 1994) and (ii) cannot
be considered as a random walk, since it is not a pure Markovian
process (there exist long tails in the autocorrelation function, e.g. see
Meiss et al. 1983, Meiss 1992 and references therein). Therefore, if one
wants to recover theoretically a law analogous to that of Lecar {\it et
al.} in cases where the perturbation is not strong, he has: (a) to
circumvent the problems in the convergence of the function $\chi (t)$
and (b) to take into account the non-Brownian and, possibly, non-
Markovian nature of transport in the statistical description of the
process.

In this paper we discuss both problems. We use a method for the study
of transport in Hamiltonian systems proposed by Varvoglis {\it et al.}
(1995) in order to calculate the fractality of phase space of the model
Hamiltonian

\begin{equation}
H = {1 \over 2} p_x^2 + {1 \over 2} p_y^2 + {1 \over 2} p_z^2 +
{1 \over 2} A x^2 + {1 \over 2}
 B y^2 + {1 \over 2} C z^2 - \epsilon xz^2 - \eta yz^2
\end{equation}
for the particular case $A$ = 0.9, $B$ = 0.4, $C$ = 0.225, $\epsilon$ =
.56, $\eta$ = 0.20 and $h$ = 0.00765 which has been studied extensively
by Contopoulos and Barbanis (1989, 1995) and Barbanis {\it et al.}
(1997). We conjecture that the fractality may be correlated to the
evolution of the function $\chi (t)$ and we discuss the possible
usefulness of a ``Local Lyapunov time'' of a trajectory, in accordance to
the ideas of Benkadda et al. (1994) and Morbidelli (1997).

In order to study transport in the phase space of the Hamiltonian (1) we
work as follows. We construct a six-dimensional grid in phase space and
we calculate the phase space volume, $V(t)$, of the elementary cells
explored by a trajectory up to time $t$. All numerically integrated
trajectories start on the $z = 0$ plane with an initial velocity
perpendicular to it ($p_x = p_y = 0$ and $p_z$ calculated
from the integral of energy). A typical example of the evolution of
$V(t)$ for three trajectories, a "central" one and two adjacent with $x_0$
differing by $\Delta x = \pm 10^{-7}$, is given in Fig. 1. We see that,
in accordance to its definition, $V(t)$ is the superposition of "step
functions" and we note that these "steps" seem to have sizes of widely
different values. We interpret the form of $V(t)$ as follows. Consecutive
small steps indicate that the trajectory is diffusing slowly, according to
the "brownian" model of random walk, in a phase space region
characterised by common properties and, in particular, by a common
rate of separation, implying a well defined LCN. A "large" step indicates
that the trajectory has moved to an adjacent phase space region with
different properties.
\begin{figure}
\resizebox{\hsize}{!}{\rotatebox{0}{\includegraphics*{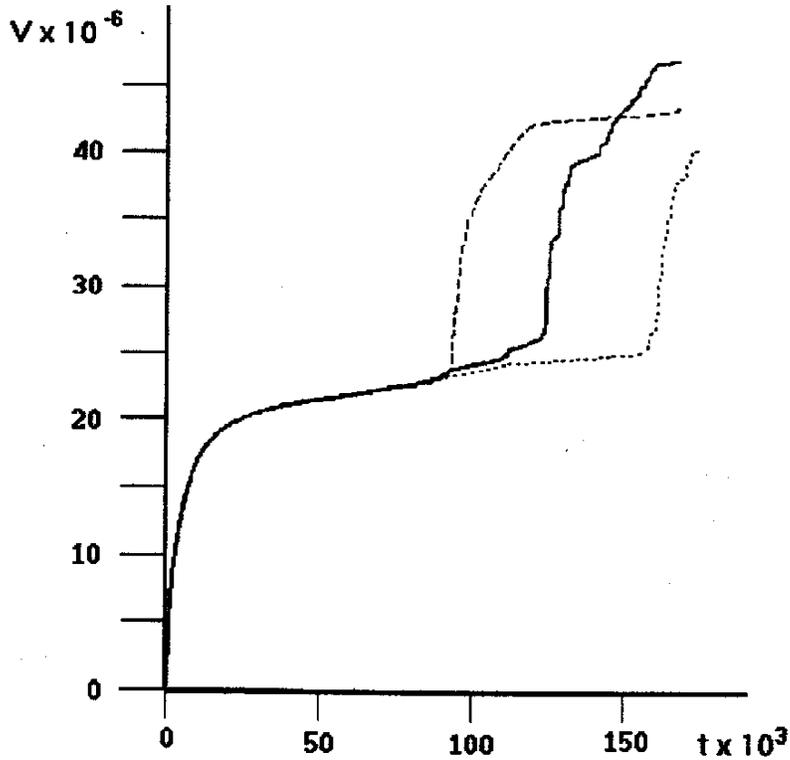}}}
\caption{The function $V(t)$ for three trajectories with initial
conditions $y_0$ = 0.032 and $x_0$ = 0.01725 (dotted curve), $x_0$ =
0.0172499 (solid curve) and $x_0$ = 0.0172501 (dashed curve).}
\end{figure}

\begin{figure}
\resizebox{\hsize}{!}{\rotatebox{0}{\includegraphics*{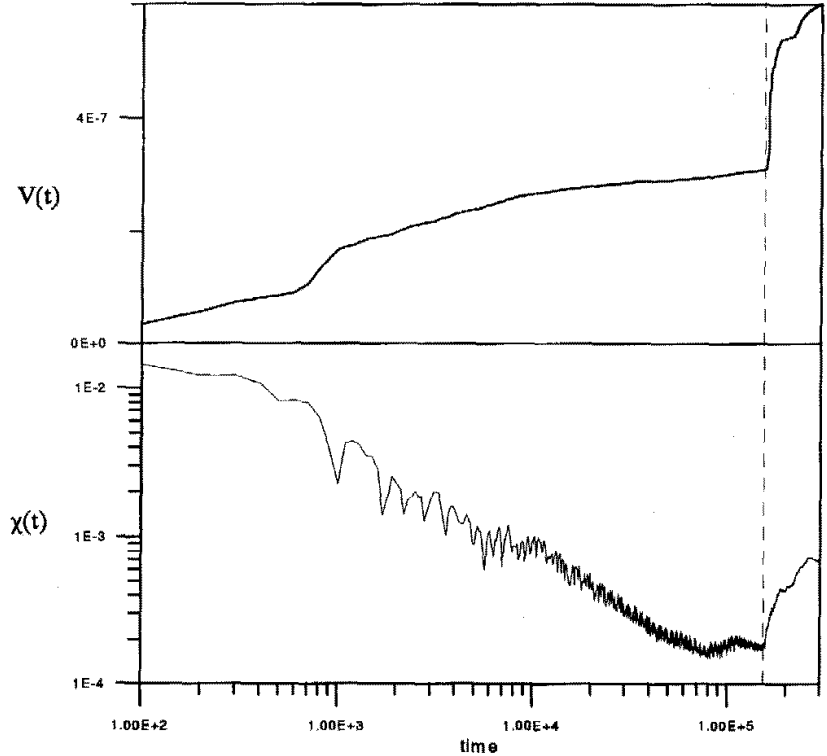}}}
\caption{Plots of $V(t)$ and $\chi (t)$ for the "central" trajectory of
Fig. 1 (dotted curve).}
\end{figure}
\section{Results}

We test this interpretation by plotting simultaneously the functions
$V(t)$ and $\chi (t)$. A typical example of our results is presented in
Fig. 2, where we have plotted the above functions for the "central"
trajectory of Fig. 1. We see that, up to $t_1 \approx 1 \cdot 10^5$, the
trajectory moves in a phase space region whose volume increases slowly
with time. From this segment one would infer that the LCN is close to
zero, since the corresponding function $\chi (t)$ decays almost
exponentially. After $t_1$, however, the trajectory migrates suddenly to
a region "outside" the initial one and the function $\chi (t)$ does not
show signs of levelling off up to the end of our numerical calculations,
at $t_{fin} = 3 \cdot 10^5$. Aside from the fact that this picture
corroborates our conjecture, concerning the relation between the
transport process and the evolution of $\chi (t)$, Fig. 2 is an example of
the practical difficulties arising in the numerical calculation of LCN's:
the limiting behaviour of $\chi (t)$ is governed by the properties of the
"outside" regions of phase space and, therefore, it does not reflect the
properties of the initial region
where the trajectory started. In the next section we discuss how these
difficulties can be circumvented by the definition of a suitable ``Local
Lyapunov time''.
\begin{figure}
\resizebox{\hsize}{!}{\rotatebox{0}{\includegraphics*{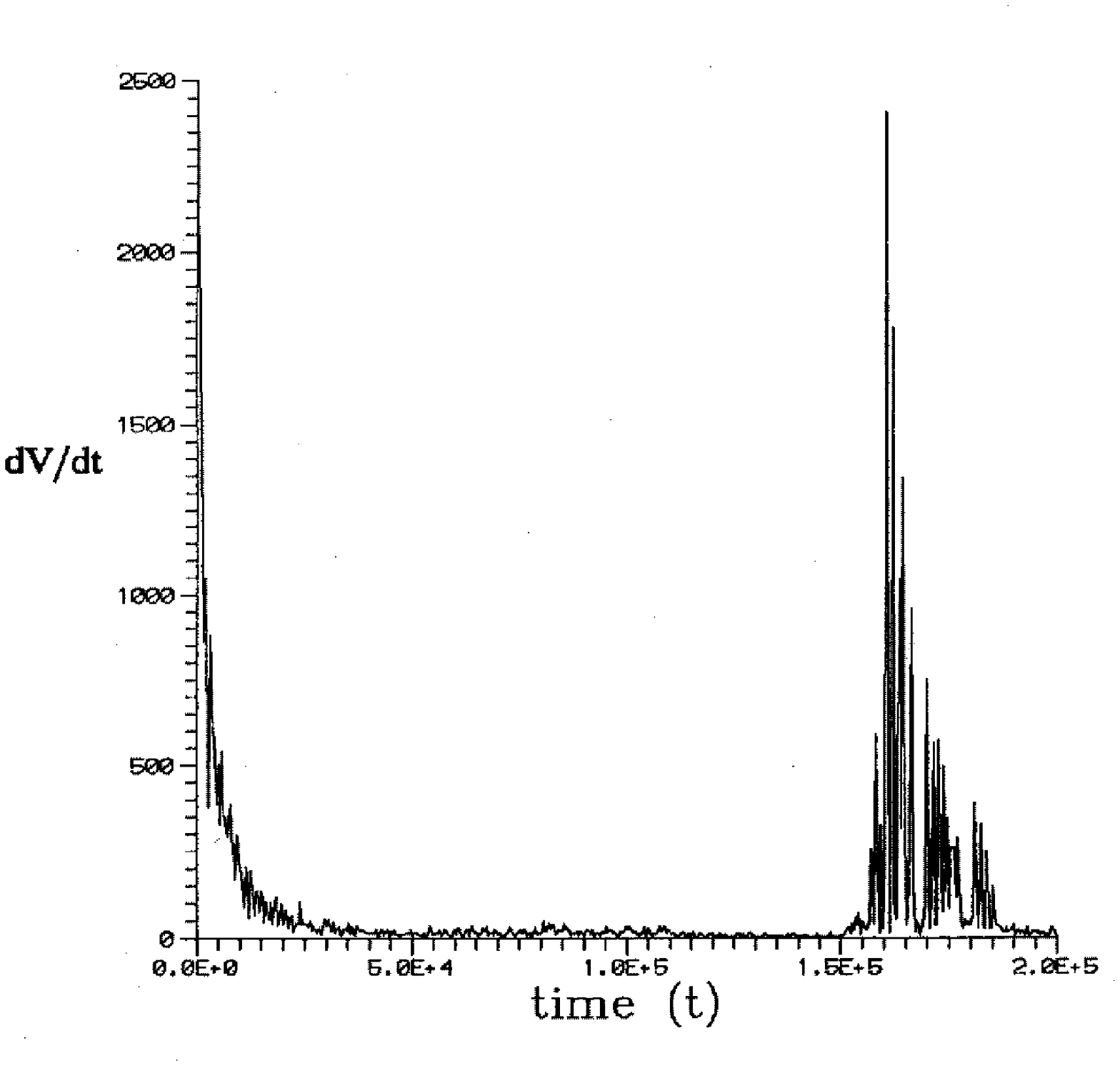}}\hspace{1cm}
\rotatebox{0}{\includegraphics*{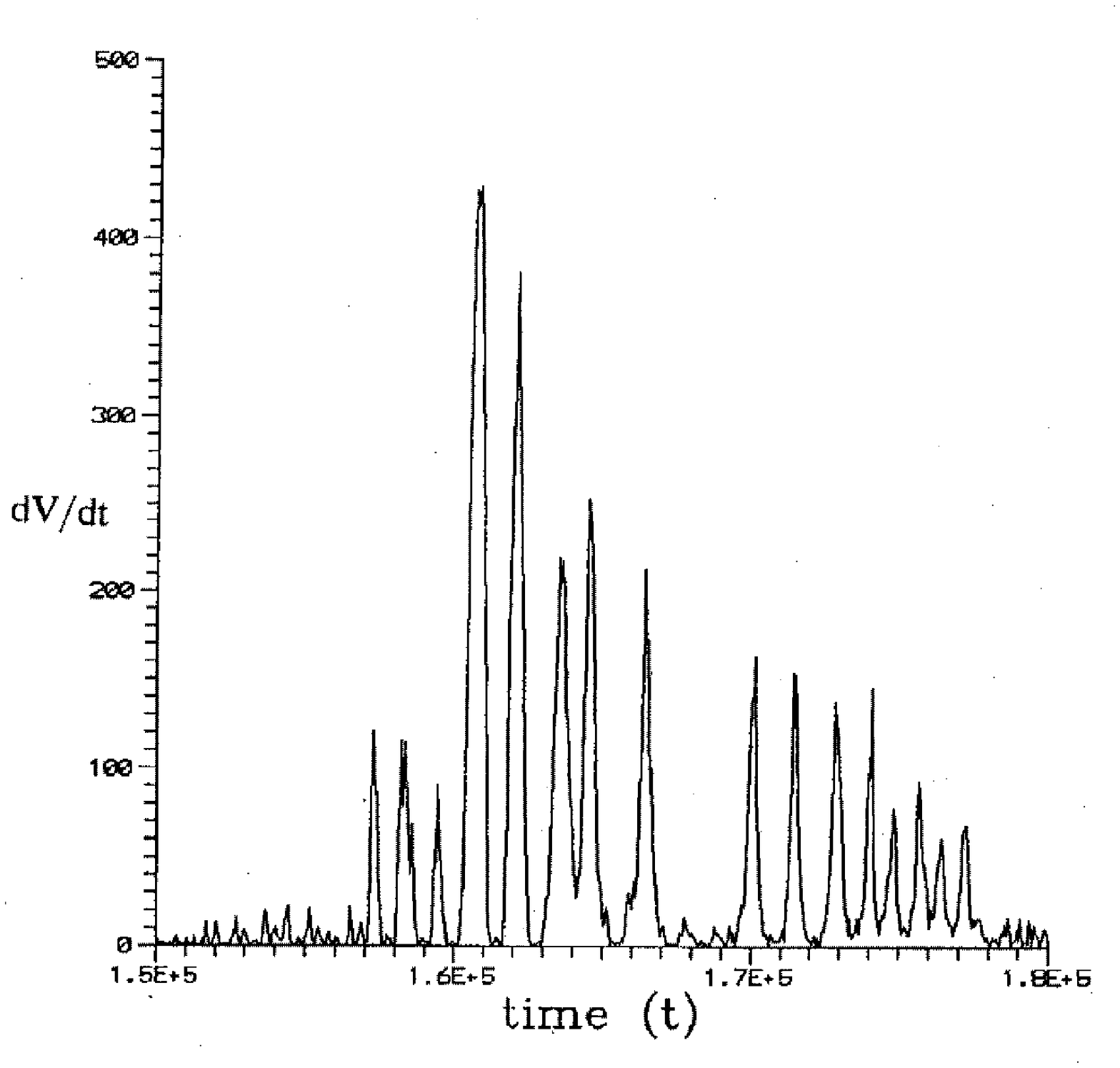}}}
\vskip 0.1cm
\resizebox{\hsize}{!}
{\hspace{3 in}\rotatebox{0}{\includegraphics*{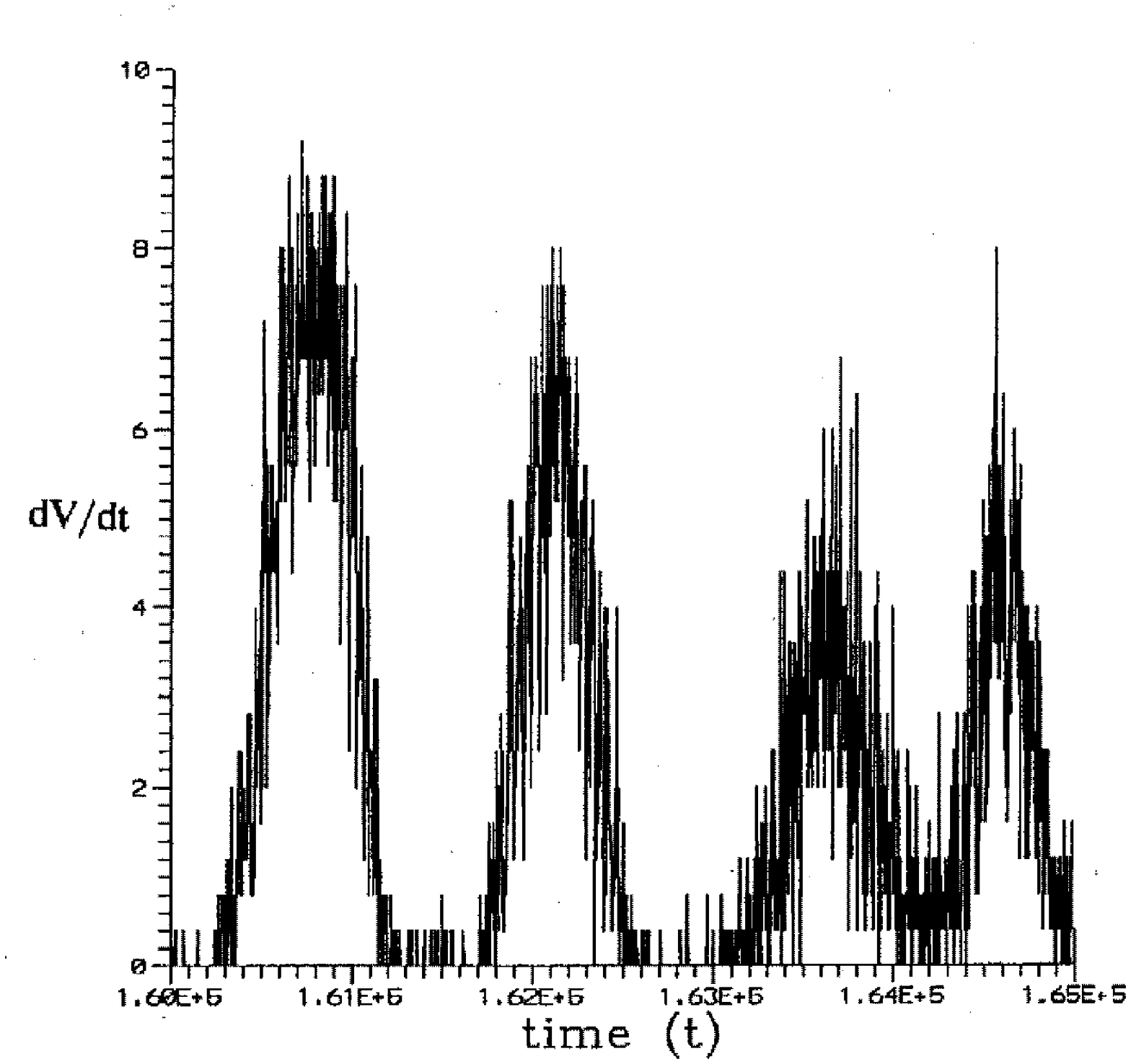}}
\hspace{3 in}} 
\caption{(3a) Plot of $dV/dt$ of the trajectory of Fig. 2. (3b) A
magnification of Fig. 3a. (3c) A magnification of Fig. (3b).}
\end{figure}

Proceeding now to the second problem mentioned, we note that, if the
phase space has a fractal structure, then the function $V(t)$
should show some sort of self-similarity. However the time series
describing numerically $V(t)$ cannot be used in a fractal analysis
algorithm, since it is not {\it stationary} (i.e. it shows long-range
changes in the mean level). Therefore we decided to use instead its
derivative, $dV/dt$, which has the above property. We see that, indeed,
$dV/dt$ consists of "spikes" of all orders, each one corresponding to a
"jump" in $V(t)$ (Figs. 3). After a transient of large spikes,
corresponding to the "spreading" of the trajectory in the initially
available phase space region, the value of the function drops until $t
\approx 1.5 \cdot 10^5$, where from the "forest" of high "spikes" it is
obvious that the trajectory migrates to a nearby phase space region. A
closer examination reveals that the spikes possess a prominent self-
similar structure: consecutive magnifications of an interval of $dV/dt$
show that each individual spike consists of higher order spikes, down to
a scale size corresponding to the 6-D volume of an elementary cell (Figs.
3b and 3c). However, apart from this qualitative approach, we can obtain
a quantitative measure of the "fractality" of the function $dV/dt$
through the calculation of its generalised dimensions, $D(q)$, q = 0, 1,
2, ...  (e.g. see Schuster, 1988; McCauley, 1995). In particular $D(2)$ is
the {\it correlation dimension}. A value of $D(2) = 1$ implies total
randomness, while a value of $D(2) = 0$ complete predictability.. It
should be noted that the canonical variables $q_i (t), p_i (t)$ (or any
function of them) are not useful for the study of phase space fractality,
since the corresponding sets are "fat fractals" (Benettin {\it et al.},
1986).
\begin{figure}
\resizebox{\hsize}{!}{\rotatebox{0}{\includegraphics*{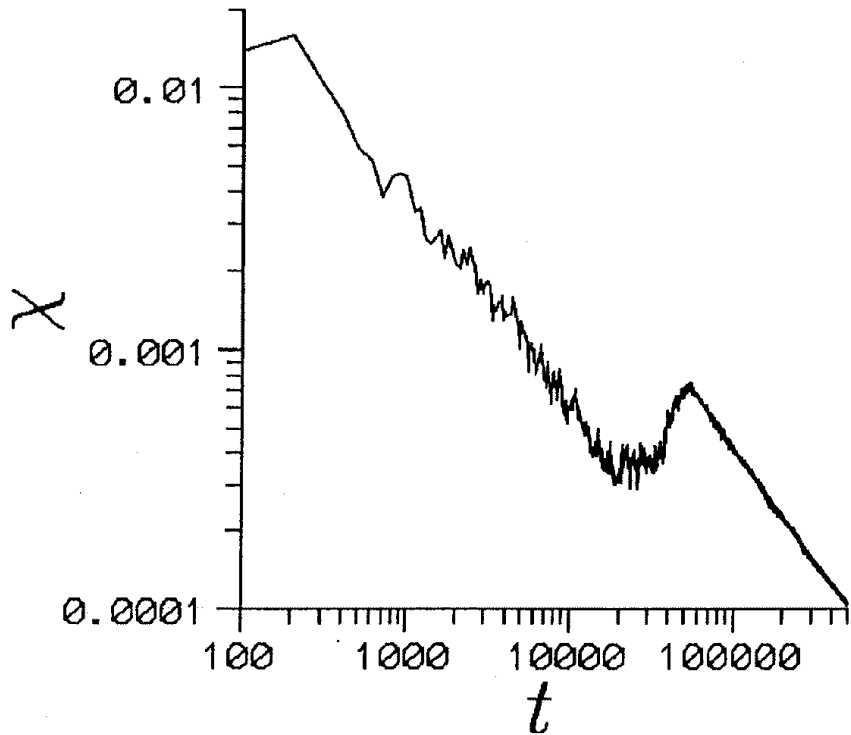}}}
\caption{Plot of $\chi (t)$ for a trajectory with initial conditions
$x_0$ = 0.0095 and $y_0$ = 0.0415.}
\end{figure}
\begin{figure}
\resizebox{\hsize}{!}{\rotatebox{0}{\includegraphics*{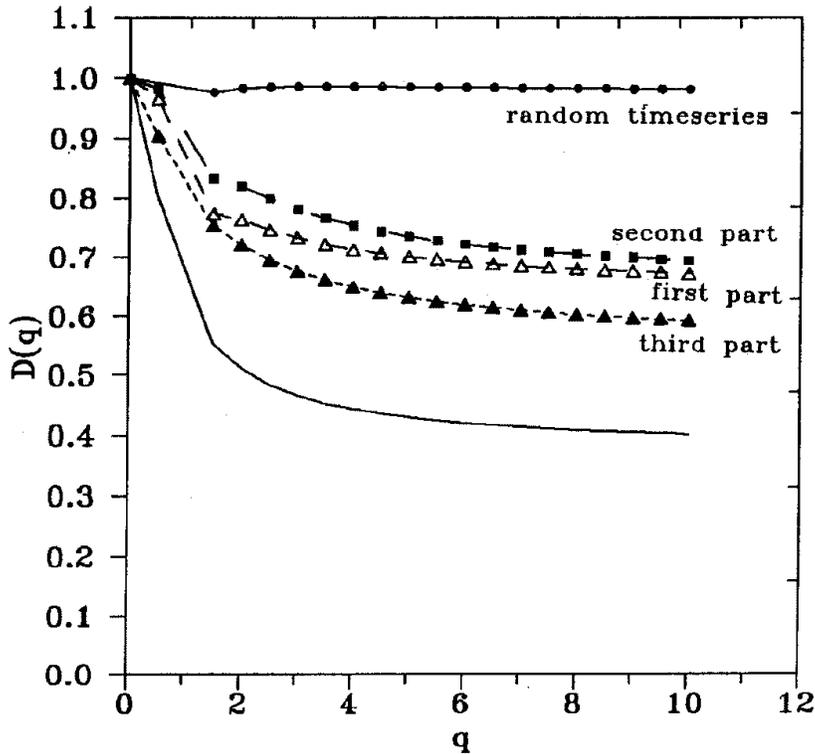}}}
\caption{The $D(q)$-spectrum of the trajectory of Fig. 4 (solid curve)
as well as that of the three segments described in the text.}
\end{figure}

From the calculation of the $D(q)$ spectrum we find that $0 < D(2) <
1$ (Fig. 5), which corroborates the results of the qualitative analysis,
showing that $dV/dt$ is a {\it multifractal} (for a definition see the
Appendix). Moreover it shows that the transport process is not
completely random. From the simultaneous plotting of $V(t)$, $dV/dt$,
$\chi (t)$ and $D(2)$ for numerous trajectories and segments of them
we found that, as a rule, $D(2)$ shows a correlation with $\chi (t)$. In
particular the segments of $dV/dt$, corresponding to time intervals
where $\chi (t)$ decays, have values of $D(2)$ lower than those of the
segments of $dV/dt$, corresponding to time intervals where $\chi (t)$
increases. On the other hand, if the trajectory is treated as a whole, the
resulting value of $D(2)$ takes an intermediate value. This fact is in
agreement with our interpretation, since, in places where the divergence
of trajectories is almost linear with time, we have weak chaos and,
therefore, a low correlation dimension. In contrast, in regions where the
divergence of trajectories in exponential, we have strong chaos and,
therefore, a high correlation dimension.

However this is not always the case. An interesting example is given in
Figs. 4 and 5. In Fig. 4 we give the function $\chi (t)$ for a trajectory
with initial conditions $x_0$ = 0.0095 and $y_0$ = 0.0415. We see that
three different trajectory segments may be defined: one for the time
interval $0 < t < 2 \cdot 10^4$, one for $2 \cdot 10^4 < t < 6 \cdot
10^4$ and one for $6 \cdot 10^4 < t < 5 \cdot 10^5$. In the first and
third segments $\chi (t)$ shows an almost exponential decay, a typical
behaviour of a weakly chaotic trajectory, while in the second segment it
shows a steep increase, characteristic of strong chaos. From what has
been said above, the trajectory has migrated from one region
characterised by an almost linear divergence of trajectories to another
one with the same property, passing briefly through a region
characterised by an exponential divergence. In Fig. 5 we give the
$D(q)$-spectra of the three segments and of the trajectory as a whole.
We see that, considering only the three segments, the "typical" relation,
discussed above, between $\chi (t)$ and $D(2)$ still holds. However in
this case the $D(2)$ of the trajectory {\it as a whole} is considerably
lower. This implies that, even while the trajectory has drifted from a
phase space region of weak chaos to another, going through one with
prominent chaotic behaviour, the third segment of the trajectory "keeps
a memory" of the first, which shows that the process in the second
segment is not Markovian!

\section{Discussion}

In a perturbed integrable Hamiltonian system we can distinguish three
regimes, according to the "amplitude" of the perturbation: those of
"small", "medium" or "large" perturbations, where the "amplitude" of the
perturbation is inferred by the dominant properties of the transport
mechanism. In the regime of small perturbation, transport is expected to
be governed by Arnold diffusion, as discussed recently by Morbidelli and
Froeschl\'e (1996). In the regime of "strong" perturbation the following
analytic result may be obtained, according to Varvoglis and Anastasiadis
(1996). Assuming that transport in a Hamiltonian system can be
described as a {\it normal} diffusion process, it can be shown that the
exit time, $T_E$, of a trajectory from a "compact" region of action space
depends on its Lyapunov time, $T_L = 1/\lambda$ (where $\lambda$
stands for the maximal Lyapunov Characteristic Number, LCN), through
a power law. Since transport in perturbed integrable Hamiltonian
systems can be modeled as normal diffusion only in regions where most
of the KAM tori are destroyed, the power law dependence appears when
the perturbation is strong.

In this paper we showed that the function $V(t)$ may be used to obtain
a quantitative measure of the fractality of phase space of a Hamiltonian
dynamical system. We discussed how the fractality of phase space is
related to the evolution of the function $\chi (t)$, in the regime of
"medium" perturbations, and the ensuing difficulties in the calculation
of LCN's. A way to circumvent these difficulties might be the definition
of a "new" quantity, related to the local (rather than the usual average)
properties of the phase space. Following the ideas of Benkadda {\it et
al.} (1994) and Morbidelli (1997), this quantity might be a ``Local
Lyapunov time'', LLT, describing strictly the "autocorrelation time", i.e.
the time interval after which a trajectory "looses memory" of its initial
conditions. This quantity may be estimated by the evolution of $V(t)$
of nearby trajectories. E.g. from Fig. 1 we can estimate that LLT
$\approx 100 \cdot 10^3$. Note that its inverse, the ``local LCN''
(LLCN), turns out to be of the order $10^{-5}$, an order of magnitude
lower than the value of LCN inferred from Fig. 2.

In order to establish a statistical relation between local properties
(i.e. rate of trajectory divergence) and global properties (transport) in
the regime of "medium" perturbations, the fractality of phase space has
to be taken into account, since it affects not only the calculation of
LCN's, but the process of transport as well. In particular the fact that
$dV/dt$ is a multifractal implies that transport in phase space
follows Levy rather than normal diffusion kinetics, while the results of
the $D(q)$ analysis show that in several cases the non-Markovian nature
of the transport process is pronounced. Therefore it is not possible to
describe transport through a classical, Fokker-Planck type, diffusion
equation and any attempt to use the "regular" random walk theory in
order to infer the statistical behaviour of trajectories in a "medium"
perturbation regime, as it is the case with the Sun-Jupiter-asteroid
problem in the ERTBP approximation, might lead to erroneous results.
It should be emphasized that, while the Fokker-Planck equation can be
appropriately generalised, in order to take into account the Levy
statistics of the random walk process (Zaslavsky, 1994), the problem of
the long-tail correlations has not been addressed in a satisfactory way up
to now. The use of the function $V(t)$, presented here, may constitute
a useful tool in the study of transport in Hamiltonian systems, in
particular when the approximation of random walk (Brownian or Levy)
may not be adequate.

\appendix

Here we give, for completeness, some basic definitions on the concepts
of fractal analysis used in the paper. The main property of a fractal set
is {\it self-similarity}, i.e. the fact that the set looks the same if one
scales appropriately the $x_i$ axes ($i=1, N$) of the space where this set
is embedded. In generalising the definition of the "dimension" of a
geometrical object, a fractal set has a {\it non-integer} dimension, which
can be understood in the following way. If for a set of points in $d$
dimensions the number $N(l)$ of $d$-spheres of diameter $l$ needed to
cover the set increases like

\begin{equation}
N(l) \propto l^{-D} ~~~~~\mbox{\rm for}~~~~ l \to 0
\end{equation}
then $D$ is the {\it fractal dimension} of the set. For a "regular"
geometrical object $D$ turns out to be an integer, which shows that the
above definition agrees with the geometrical dimension of a non-fractal
set. In the case where the scaling depends on the region of the set, the
geometrical object is a multifractal. "Naturally" occurring fractals fall in
this category.

A multifractal is characterised by more than one "fractal dimensions"
and, in this way, it contains a lot more information than a pure fractal.
These so-called {\it generalised dimensions} or $D(q)$ {\it spectrum}
can be calculated as follows. If we divide the space in cells of linear
dimension $l$ and we denote by $p_i$ the probability that a point of the
set lies in the $i-th$ box, then the generalised dimensions $D(q)$ are
given by the formula

\begin{equation}
D(q) = - \lim_{l \to 0} {1 \over {q-1}} {1 \over {\log l}} \log (\sum_
{i} p_i^q)
\end{equation}
All generalised dimensions $D(q)$ of a pure fractal set have the same
value, that of the {\it capacity dimension}, which corresponds to $q$ =
0, i.e. $D(0)$.

It is obvious that comparing two multifractals is much more difficult
than comparing two pure fractals, since it involves comparing an infinite
sequence of numbers, rather than only two (the two capacity
dimensions). In applications one usually compares only the values of
$D(2)$, since in experimental results this is the number that can be more
easily calculated. $D(2)$ has the name {\it correlation dimension}, since
it is related to the {\it correlation integral}, $C(l)$, of the time-series.
It can be proved that, in general, $D(0) > D(1) > D(2) ... > D(n)$.

\acknowledgements
 The authors would like to thank M. Georgoulis
for his help in the calculation of the $D(q)$ spectra.\par

\end{document}